\documentclass{llncs}
\usepackage{amsmath,amssymb}
\usepackage{stmaryrd} 
\usepackage{mathabx} 
\allowdisplaybreaks[3]
\usepackage{cancel}
\usepackage{tikz}
\input xy 
\xyoption{all}

\usepackage{amsmath}

\usepackage{mathtools}

\newcommand{\REL}{\REL}

\def\PL{\mathit{PL}}
\def\REL{\mathit{REL}}

\def\H2PL{\mathcal{H}^2\PL}

\def\rcb#1#2#3#4{\def\nothing{}\def\range{#3}\mathopen{\langle}#1 \ #2 \ \ifx\range\nothing::\else: \ #3 :\fi \ #4\mathclose{\rangle}}

\DeclareMathAlphabet{\mathbb}{U}{msb}{m}{n}
\DeclareSymbolFont{ams}{U}{msa}{m}{n}
\DeclareSymbolFontAlphabet{\mathams}{ams}
\DeclareMathSymbol{\filter}{\mathams}{ams}{22}

\usepackage{float} 
\usepackage{subfig}
\usepackage{relsize}
\usepackage{xspace}

\usepackage[bookmarks,unicode
           ,colorlinks=true
           ,citecolor=green!40!black
           ,urlcolor=green!40!black
           ,linkcolor=green!40!black]{hyperref}%
\usepackage[nameinlink,capitalise]{cleveref} 
\usepackage{eurosym}

\usepackage{tikz-network}
\usepackage{xurl}
\usetikzlibrary{automata,positioning,calc,shapes.multipart,arrows.meta}
\tikzstyle{obj} =[circle, minimum width=0.5cm, minimum height=0.5cm, draw=black]
\tikzstyle{aon}=[line width=0.6pt,>={Triangle[scale width=0.8] Triangle[scale width=0.8]},->]
\tikzstyle{aoff}=[line width=0.6pt,>={Rays[]},->]

\newcommand{\arrowG}[1][]{%
  \raisebox{1pt}{\tikz \draw [-{Triangle[scale width=0.8]}] (0,0) -- node[above,font=\scriptsize]{#1} (3.5mm,0);}}
\newcommand{\arrowOn}[1][]{%
  \raisebox{1pt}{\tikz \draw [aon] (0,0) -- node[above,font=\scriptsize]{#1} (3.5mm,0);}}
\newcommand{\arrowOff}[1][]{
  \tikz \draw [aoff] (0,0) -- node[above,font=\scriptsize]{#1} (3.5mm,0);}

\def\topfigrule{\kern 7.8pt \hrule width\textwidth\kern -8.2pt\relax}
\def\dblfigrule{\kern 7.8pt \hrule width\textwidth\kern -8.2pt\relax}
\def\botfigrule{\kern -7.8pt \hrule width\textwidth\kern 8.2pt\relax}

\newcommand{\mar}{\textrm{reactive graph}\xspace}
\newcommand{\mars}{\textrm{reactive graphs}\xspace}

\renewcommand{\Mars}{\textrm{Reactive graphs}\xspace}

\newcommand{\mi}[1]{\ensuremath{\mathit{#1}}\xspace}
\newcommand{\tpl}[1]{\langle #1 \rangle}
\newcommand{\myparagraph}[1]{\medskip\noindent\textbf{#1} }

\newcommand{\Marge}{\textcolor{green!40!black}{\textsf{Marge}}\xspace}
\newcommand{\urlpp}[1]{{\small \url{#1}}\xspace}
\newcommand{\urlspp}[1]{{\footnotesize \url{#1}}\xspace}
\newcommand{\action}[1]{\bgtext{lightgray!50!white}{#1}\xspace}

\usepackage[normalem]{ulem}
\usepackage[textwidth=39.0mm,textsize=scriptsize]{todonotes}
\newcommand{\bgtext}[1]{%
  \bgroup\markoverwith {\textcolor{#1}{\rule[-0.5ex]{2pt}{11pt}}}\ULon}

\newcommand{\dwavy}[1][\Gamma^{\oplus},\Gamma^{\ominus}]{\mathrel{\dwavyAux\!\!\dwavyAux\!_{#1}}}
\newcommand{\dwavyAux}{\resizebox*{!}{2.5mm}{\rotatebox{75}{\!{\footnotesize\uwave{\hspace{4mm}}}\!}}}
\newcommand{\dline}[1][\Gamma^{\oplus},\Gamma^{\ominus}]{\mathrel{\sslash\!_{#1}}}

\begin{document}
\title{Reactive graphs in action
\\{\normalsize (extended version)}
\thanks{This work extends a FACS 2024 publication~\cite{marge-facs} and includes minor corrections.}
}

\author{David Tinoco\inst{1} \and
Alexandre Madeira\inst{1} \and \\
Manuel A. Martins\inst{1} \and José Proença\inst{2}}
\institute{CIDMA, Dep. Mathematics, University of Aveiro, Aveiro, Portugal
\and
                 CISTER, Faculty of Sciences, University of Porto, Porto, Portugal 
}

\maketitle     
\begin{abstract}
\emph{Reactive graphs} are transition structures whereas edges become active and inactive during its evolution, that were introduced by Dov Gabbay from a mathematical's perspective.
This paper presents \Marge
(\urlspp{https://fm-dcc.github.io/MARGe}), a web-based tool to visualise and analyse reactive graphs enriched with labels. \Marge animates the operational semantics of reactive graphs and offers different graphical views to provide insights over concrete systems. We motivate the applicability of reactive graphs for adaptive systems and for featured transition systems, using \Marge to tighten the gap between the existing theoretical models and their usage to analyse concrete systems.

\end{abstract}
\section{Introduction and motivation}
A reactive graph is a transition structure that updates its transitions along its execution. This concept has been introduced by Dov Gabbay in \cite{GabaySeminal}. It generalizes the static notion of a graph by incorporating high-order edges that capture updates on the accessibility relations.
The notion of reactivity for these structures is not coined only in the standard sense of Harel and Pnueli~\cite{pnueli}, as systems that react to their environment and are not meant to terminate, but as systems whose accessibility relation is 
a result of the transformations induced by the transitions executed so far.
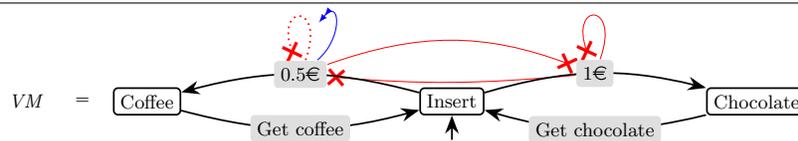
\begin{figure}[b]
    \centering
  \resizebox{11cm}{!}{
    {

\begin{tikzpicture}[node distance={15mm}, thick, main/.style = {draw,rectangle, rounded corners = 2pt}, action/.style = {rectangle, fill = lightgray!50!white, text = black, rounded corners = 2pt}, minimum size = 0.3cm,>={Stealth[length=3mm]}] 
    
    \tikzstyle{enb}=[>={latex[length=3mm]}]

    \begin{scope}
        \node[main] (insert) at (0,0){Insert};
        \node[main] (cofee) at (-5,0){Coffee};
        \node[main] (choco) at (5,0){Chocolate};

        \draw[->,bend left=15]  (insert)  to node[action](1choco){$1$\euro} (choco);
        \draw[<-,bend right=15] (insert)  to node[action](bot){Get chocolate} (choco);
        
        \draw[->,bend right=15] (insert)  to node[action](5c){$0.5$\euro} (cofee);
        \draw[<-,bend left=15] (insert)  to node[action]{Get coffee} (cofee);

    \end{scope}

    \begin{scope}[on background layer]
        \draw[color = red,-{Rays[length=10pt,width=10pt,line width= 1.5pt]},out= 70, in = 110,looseness = 15] (1choco) to node(top){} (1choco);
        \draw[color = red,-{Rays[length=10pt,width=10pt,line width= 1.5pt]},bend left=5] (1choco) to (5c);
        \draw[color = red,-{Rays[length=10pt,width=10pt,line width= 1.5pt]},bend left=20] (5c) to (1choco);

        \draw[color = red,-{Rays[length=10pt,width=10pt,line width= 1.5pt]},out= 70, in = 110,looseness = 15, dotted, thick] (5c) to node[pos=0.2](0){} (5c); 
        \draw[color = blue,enb,->>,
              out= 40, in = 50,looseness = 2.3] (5c) to (0);

        \coordinate[yshift=-4mm](start)at(insert.south);
        \draw[->,thick](start)to(insert);

    \end{scope}

    \pgfresetboundingbox
    \node[inner sep=0]at(cofee.west){};
    \node[inner sep=0]at(choco.east){};
    \node[inner sep=0,yshift=-3mm]at(top){};
    \node[inner sep=0]at(bot.south){};

    \node[xshift=-10mm,left](vm) at (cofee.west){$\mathit{VM}$};
    \node at ($(vm.east)!0.5!(cofee.west)$) {$=$};
\end{tikzpicture} }}
    \caption{A reactive graph of a vending machine offering two different products}
    \label{fig:vd1}
\end{figure}

Let us consider the model of a simple vending machine in \cref{fig:vd1} to motivate reactive graphs. 
Edges in reactive graphs can be active or inactive, and only transitions involving active edges can be executed.
All edges in our vending machine are active, except one with a dotted line.
Executing a transition triggers an update on the set of active and inactive edges.

The machine \mi{VM} in \cref{fig:vd1} can receive from the user at most 1\euro. The arrows between states represent \emph{ground edges}, which are labelled with actions; the others arrows represent \emph{hyper edges}, i.e., edges that can activate ($\arrowOn$) and deactivate ($\arrowOff$) edges. 
When the action \action{1\euro} is performed some edges are deactivated and the machine goes to the \mi{Chocolate} state. More specifically, both edges labelled by \action{0.5\euro} and \action{1\euro} are deactivated. Executing instead the edge labelled by \action{0.5\euro} would enable a deactivating edge, and executing it a second time would deactivate it.

A solid supporting theory for these models, including proposals of specification logics, has been studied in the last years (e.g. \cite{GabbayM09,GabbayM12}) and was summed up in the book \cite{GabbayBook}.
The main advantage of reactive graphs with respect to traditional labelled transition systems (LTS) is the compact representation of dynamic systems.
Gabbay showed the encoded LTS of a given reactive system can have an exponentially larger number of states~\cite[Prop. 8.8]{GabbayBook}.
For example, our vending machine can be expressed with an LTS with seven states, depicted in \cref{fig:vd2}. In a larger example borrowed from Cordy et al.~\cite[Fig.~1]{cordy2013model}, presented in \cref{appendix: FTS}, the encoded LTS has 7x more states and 4.6x more edges (including hyper edges).
Furthermore, we believe that many examples become easier to understand and to maintain with reactive graphs than with traditional LTS.
\begin{figure}[t]
    \centering
   \resizebox{10cm}{!}{

\begin{tikzpicture}[node distance={15mm}, thick, main/.style = {draw,rectangle, rounded corners = 2pt}, action/.style = {rectangle, fill = lightgray!50!white, text = black, rounded corners = 2pt}, minimum size = 0.3cm,>={Stealth[length=3mm]}] 
    
    \begin{scope}
        \node[main] (inserti)  at (0,0){Insert};
        \node[main] (choco)    at (3, 0.7){Chocolate};
        \node[main] (insertch) at (8, 0.7){Insert};
        \node[main] (cofee1)   at (3,-0.7){Coffee};
        \node[main] (insertc1) at (6,-0.7){Insert};
        \node[main] (cofee2)   at (8.5,-0.7){Coffee};
        \node[main] (insertc2) at (11.5,-0.7){Insert};

        \draw[->,bend left=15]  (inserti)  to node[action]{$1$\euro} (choco);
        \draw[->] (choco)  to node[action]{Get chocolate} (insertch);
        
        \draw[->,bend right=15]  (inserti)  to node[action]{$0.5$\euro} (cofee1);
        \draw[->] (cofee1)  to node[action,rotate=45]{Get coffee} (insertc1);
        \draw[->] (insertc1)  to node[action,rotate = 45]{$0.5$\euro} (cofee2);
        \draw[->] (cofee2)  to node[action,rotate =45]{Get coffee} (insertc2);


        \coordinate[xshift=-4mm](start)at(inserti.west);
        \draw[->,thick](start)to(inserti);

    \end{scope}

\end{tikzpicture} }
    \caption{The LTS of the vending machine in \cref{fig:vd1}}
    \label{fig:vd2}
\end{figure}
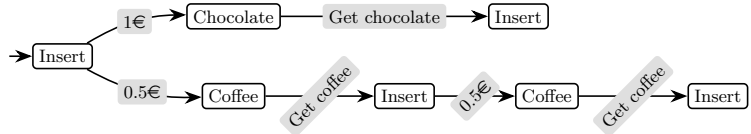

\noindent {\bf Reconfigurable systems and variability.}
Not all scenarios and application domains can exploit the advantages of modelling by reactive graphs. The benefits of compactness and being more intuitive are more evident when analysing \textbf{reconfigurable systems}. These are systems that operate in different modes of execution, e.g., an operating systems that supports users with different permissions or different performance modes (e.g. an ocean exploring robot that adjusts its behaviour based on the distance of its base, its energy levels, and its environmental conditions).

Work on \emph{software product lines} (SPL) focuses mainly on \textbf{configurable systems}, i.e., how to develop, maintain, and reason over families of software that share many commonalities. Feature transition systems ~\cite{FTSterBeek,cordy2013model} are structures often used to model such systems, which are enriched with annotations over features, allowing an initial configuration of the variant to select which transitions are active. A branch of SPL focuses on \emph{dynamic SPL}~\cite{CBTRH14}
addressing \emph{reconfigurable and self-adaptive systems}, in which the configuration can change over time, which relates to our reactive graphs.

This paper introduces \Marge:
an open-source web-based prototype tool designed for the visualisation and analysis of labelled reactive graphs. It includes several examples, both from the literature on reactive graphs~\cite{GabbayBook} and from work on dynamic SPL~\cite{cordy2013model}. The goal of \Marge is to help increasing the adoption of reactive graphs,
providing insights over the capabilities and challenges of modelling reconfigurable systems with reactive graphs,
and to expose these to different domains to increase it applicability. Currently \Marge does not aim at contributing directly to the community on adaptative SPLs, since it still misses a more user-friendly specification language and mechanisms to support larger systems.
\Marge provides support to:
(1) \emph{visualise} reactive graphs,
(2) \emph{animate} its operational semantics,
(3) \emph{explore} the full state-space of its underlying LTS,
(4) \emph{verify} properties such as deadlocks and conflicts, and
(5) \emph{compare} reactive graphs using an observational equivalence.

\section{Multi-actions reactive graphs}\label{sec:defns}
A \emph{multi-action reactive graph}, or simply \emph{reactive graph}, is a labelled transition system with transitions enriched with a reaction that activates and deactivates transitions, defined formally below.

\begin{definition}[Reactive graph]\label{def:Mar}
  A \emph{Multi-Action Reactive Graph} is a tuple
    $M = (W, Act, E, \arrowG, \arrowOn, \arrowOff,\overline{\cdot},w_0,\alpha_0)$ where:
  \begin{itemize}
    \item $W\neq \emptyset$ is the \emph{set of states};
    $Act$ is the \emph{set of actions};
    $E$ is the \emph{set of edges};
    \item $\arrowG ~\subseteq~ W \times Act \times W$ is the set of ground edges;
     $\arrowOn  ~\subseteq~ E\times E$ is the set of activating edges;
     $\arrowOff ~\subseteq~ E\times E$ is the set of deactivating edges;
     $\overline{\cdot}~:\, E \longrightarrow (\arrowG\cup\arrowOn\cup\arrowOff)$ is an injective function that maps edges in $E$ to their internal details;
    \item $w_0 \in W$ is the \emph{initial state};
     $\alpha_0 \subseteq E$ is the set of \emph{initially active edges}.
\end{itemize}
\end{definition}

\noindent\textbf{Notation.}
  Recall the vending machine in \cref{fig:vd1}.
  When formalising this as a \mar we say that:
  (i) \mi{Coffee} belongs to $W$ (among others),
  (ii) \bgtext{lightgray!50!white}{\,0.5\euro\,} is an action in \mi{Act},
  $\tpl{\mi{Insert},\text{\action{0.5\euro}},\allowbreak\mi{Coffee}}$ is a ground edge in $\arrowG$,
  (iii) $\tpl{e^\dag,e^\dag}$ is a deactivating edge in $\arrowOff$ where $\overline{e^{\dag}} = \tpl{\mi{Insert},\text{\action{0.5\euro}},\allowbreak\mi{Coffee}}$, 
  (iv) $w_0 = \mi{Insert}$,
  and
  (v) $\alpha_0$ is the set of edges in $E$ without the deactivating edge $\tpl{e^\dag,e^\dag}$. 

\medskip

A \mar has an initial state $w_0$ and an initial set of active edges $\alpha_0$. Evolving a \mar means transitioning to a new state, connected by an enabled ground edge from $w_0$, and updating the set of active edges. We start by formalising the set of activate and deactivate edges by another given edge, and then formalise the evolution of \mars.

\begin{definition}[Activation and deactivation]
    Given a \mar $M$, an edge $e\in E_M$ and a set of active edges $\alpha \subseteq E_M$, we define the set of edges activated by $e$ (resp. deactivated by $e$), written $\mathsf{on}(e,\alpha)$ (resp. $\mathsf{off}(e,\alpha)$) as follows. 
  \begin{align*}
    \mathsf{from}(e_s) =& \{e ~|~ \exists e_t \cdot \overline{e} = (e_s,e_t)\}
   \\
    \mathsf{from\mkern-2mu}^{*}(e,\alpha) = &
      \begin{array}{l}
        \bigcup_{r\in (\mathsf{from}(e) \cap \alpha)}
        \mathsf{from\mkern-2mu}^{*}(r,\alpha\backslash \{e\})\cup \{r\}
      \end{array}
    \\
    \mathsf{on}(e,\alpha)=&
      \{e_t ~|~ 
        e_{\mi{trg}} \in \mathsf{from\mkern-2mu}^{*}(e,\alpha) 
        \land
        \exists e_s \cdot \overline{e_{\mi{trg}}} = (e_s,e_t) \in \arrowOn
      \}  
    \\
    \mathsf{off}(e,\alpha)=& 
      \{e_t ~|~ 
        e_{\mi{trg}} \in \mathsf{from\mkern-2mu}^{*}(e,\alpha) 
        \land
        \exists e_s \cdot \overline{e_{\mi{trg}}} = (e_s,e_t) \in \arrowOff
      \}  
  \end{align*}
\end{definition}

Intuitively $\mathsf{from}(e_s)$ returns the hyper edges that start from $e_s$,
$\mathsf{from\mkern-2mu}^{*}$ keeps traversing $\mathsf{from}$ to collect all (active) hyper edges triggered from a single edge,
and
$ \mathsf{on}(e,\alpha)$ (resp. $ \mathsf{off}(e,\alpha)$) collect all the targets triggered from $e$ by an activating (resp. deactivating) edge.
For example, in the vending machine in \cref{fig:vd1} we have that $\mathsf{off}(e_1,E) = \{e_1,e_2\}$, where 
  $\overline{e_1}=\tpl{\mi{Insert},\text{\action{1\euro}},\mi{Chocolate}}$ and
  $\overline{e_2}=\tpl{\mi{Insert},\text{\action{0.5\euro}},\mi{Coffee}}$.
This means that executing the edge from \mi{Insert} to \mi{Chocolate} triggers the deactivating edges $e_1$ and $e_2$.
Using this notion of (de)activation, the evolution of a \mar is formalised below.

\begin{definition}[Semantics]\label{def:sem}
  The semantics of a \mar $M$ is given by the evolution of a configuration $\tpl{w,\alpha}$ of a state $w\in W$ and active edges $\alpha\subseteq E$, starting from the initial configuration $\tpl{w_0,\alpha_0}$, given by the rule below.
  \\[1mm] \centerline{$\begin{array}{c}
      \exists e\in \alpha ~\cdot~
        \overline{e} = \tpl{w,a,w'} ~\land~
        \alpha' = \big( \alpha \cup \mathsf{on}(e,\alpha) \big)
                  \backslash \mathsf{off}(e,\alpha)
      \\\hline
      \tpl{w,\alpha} \xrightarrow{a}_M  \tpl{w',\alpha'}
  \end{array}$}
  \end{definition}

Using the semantics above to our reactive vending machine from \cref{fig:vd1} we obtain the LTS depicted in \cref{fig:vd2}.
This semantics differs from Gabbay's semantics by atomically collecting all activate edges before applying their (de)activations effects, instead of activating and deactivating edges during the traversal of triggered edges. It also introduces a bias: whenever an edge is both activated and deactivated in a step, deactivation takes precedence. However this may not be intended, which we will address in the next section over contradictory effects.

\medskip

\noindent {\bf Relevant properties of reactive graph.}\label{sec:ptroperties}
As seen in \cref{def:sem}, the behaviour of a \mar $M = (W, Act, E, \arrowG,$ $  \arrowOn, \arrowOff,\overline{\cdot},w_0,\alpha_0)$ from a configuration $\tpl{w,\alpha}$ can be represented by the LTS induced by relation $\to_M = \bigcup\{ \xrightarrow{a}_M \mid a\in Act\}$.
Many standard properties of \mars can be defined over the LTS induced by the semantics of \mar, 
namely:

\smallskip
\noindent {\it Deadlocks.} 
A deadlock is a state from which there is no transition (in our case an \emph{active transition}), often undesirable.
In \mars we can also search for deadlocks
by traversing the induced LTSs from $w_0$ while searching for states without outgoing transitions.

\smallskip
\noindent   {\it Unreachable states.} 
An unreachable state, also undesirable in many systems, is a state that cannot be reached from the initial configuration.
    
\smallskip
\noindent  {\it Observational equivalence.}
    As in standard LTS, two configurations are said to be equivalent if they behave in the same way. One way of defining such kind of equivalences is by means of their induced LTS: two configurations are behavioural equivalent if their induced LTS are bisimilar.
  
\smallskip
Other properties that can be analysed directly over \mars include:

\smallskip
\noindent {\it Contradictory effects.}
A contradictory effect is when a step triggers both the activation and deactivation of the same edge.
 The semantics in \cref{def:sem} gives priority to disabling, but often these situations are the result of bad design decisions that should be avoided, and can be signalled as warnings.

\smallskip
\noindent {\it Unreachable transitions.}
Similarly to unreachable states, (hyper) edges that cannot be fired are usually undesirable or a result of a bad understanding of a system. Hence it is a property that can also be investigated directly over \mars.

\smallskip

\noindent
{\bf Products on reactive graphs.}\label{sec: products}
Synchronous and asynchronous products of RG can be defined in the standard way. This section discusses
a new product, called \emph{intrusive product}. It allows the connections between two RGs, i.e., the execution of an action in a given machine can interfere with the activation/deactivation actions of the other machine, and vice-versa.

\begin{definition}
    Given two multi-action reactive graphs $M_1, M_2$, and $\Gamma^{\oplus},\Gamma^{\ominus} \subseteq E_1 \times E_2 \cup E_2 \times E_1$ is the set of intrusive edges between $M_1$ and $M_2$. The effects produced by $e \in E_{M_i}$ in $M_i$ is given for the set follow:
\\[1mm]
  \centerline{
    $\alpha_i(\Gamma^{\oplus},\Gamma^{\ominus},e) =
      \big(\alpha_i \cup \mathsf{on}(e,\alpha_i) \cup \Gamma^{\oplus}(e) \big)
               \setminus
      \big(\mathsf{off}(e,\alpha_i) \cup \Gamma^{\ominus}(e) \big)$}
    \end{definition}

\newcommand{\mis}[1]{{\small\mathit{#1}}}

\begin{figure}[t]
    \centering
    \resizebox{9cm}{!}{

\begin{tikzpicture}[node distance={15mm}, thick, main/.style = {draw,rectangle, rounded corners = 2pt}, action/.style = {rectangle, fill = lightgray!50!white, text = black, rounded corners = 2pt}, minimum size = 0.3cm,>={Stealth[length=3mm]}] 
    
    \tikzstyle{enb}=[>={latex[length=3mm]}]

        \node[main] (insert) at (0,0){Insert};
        \node[main] (cofee) at (-4,0){Coffee};
        \node[main] (choco) at (4,0){Chocolate};

        \draw[->,bend left=15]  (insert)  to node[action](1choco){$1$\euro} (choco);
        \draw[<-,bend right=15] (insert)  to node[action]{Get chocolate} (choco);
        
        \draw[->,bend right=15] (insert)  to node[action](5c){$0.5$\euro} (cofee);
        \draw[<-,bend left=15] (insert)  to node[action]{Get coffee} (cofee);


        \draw[color = red,-{Rays[length=10pt,width=10pt,line width= 1.5pt]},out= 70, in = 110,looseness = 15] (1choco) to (1choco);
        \draw[color = red,-{Rays[length=10pt,width=10pt,line width= 1.5pt]},bend left=5] (1choco) to (5c);
        \draw[color = red,-{Rays[length=10pt,width=10pt,line width= 1.5pt]},bend left=20] (5c) to (1choco);

        \draw[color = red,-{Rays[length=10pt,width=10pt,line width= 1.5pt]},out= 70, in = 110,looseness = 15, dotted, thick] (5c) to node[pos=0.2](0){} (5c); 
        \draw[color = blue,enb,->>,
              out= 40, in = 50,looseness = 2.3] (5c) to (0);

        \coordinate[yshift=-4mm](start)at(insert.south);
        \draw[->,thick](start)to(insert);



        \node[main] (user) at (-2,2.2){User};
        \node[main] (select) at (2,2.2){Select};
        \coordinate[xshift=-4mm](start2)at(user.west);
        \draw[->,thick](start2)to(user);

        \draw[->,bend left=15]  (user)  to node[action](coin){Coin} (select);
        \draw[<-,bend right=15] (user)  to node[action](getproduct){Get product}    (select);

        \draw[color = red,-{Rays[length=10pt,width=10pt,line width= 1.5pt]},out= 30, in = -90] (5c) to node[pos=0.2](0){} (getproduct);

\end{tikzpicture} }
    \caption{An example of two reactive graphs and one intrusive edge}
    \label{fig:VM_U}
\end{figure}
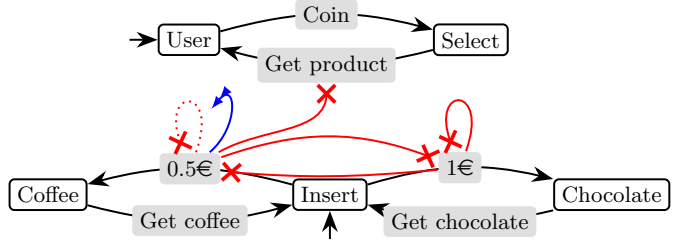

 \Cref{fig:VM_U} illustrates an 
 intrusive product of $\mi{Usr}\dline[\emptyset,\Gamma^{\ominus}] \mi{VM}$, where \mi{Usr} is the upper RG and
  $\Gamma^{\ominus} = \{
         \tpl{\tpl{\mis{Insert},\text{\action{0.5\euro}},\allowbreak\mis{Coffee}},\tpl{\mis{User},\text{\action{Get product}},\allowbreak\mis{Select}}}
   \}$.

Formally the asynchronous product ($\dline$) is defined by the rules below.
\\[0mm]
\begin{equation*}\label{RO: LRG_PIA1}
      \begin{array}{c}
            \exists ~ e \in \alpha_1 \cdot \overline{e} = s_1 \xrightarrow{a} s_1'   \land         \alpha_1' = \big( \alpha_1 \cup \mathsf{on}(e,\alpha_1) \big)
                  \backslash \mathsf{off}(e,\alpha_1)  \land \alpha_2' = \alpha_2(\Gamma^{\oplus},\Gamma^{\ominus},e)
            \\\hline
            \tpl{s_1,\alpha_1} \dline \tpl{s_2,\alpha_2} ~ \xrightarrow{a}  ~  \tpl{s'_1,\alpha_1'} \dline ,\tpl{s_2,\alpha_2'}
      \end{array}
\end{equation*}
\vspace{-0.2cm}
\begin{equation*}\label{RO: LRG_PIA2}
      \begin{array}{c}
            \exists ~ e \in \alpha_2 \cdot \overline{e} = s_2 \xrightarrow{a} s_2'   \land         \alpha_2' = \big( \alpha_2 \cup \mathsf{on}(e,\alpha_2) \big)
                  \backslash \mathsf{off}(e,\alpha_2)  \land \alpha_1' = \alpha_1(\Gamma^{\oplus},\Gamma^{\ominus},e)
            \\\hline
            \tpl{s_1,\alpha_1} \dline \tpl{s_2,\alpha_2} ~ \xrightarrow{a}  ~ 
            \tpl{s_1,\alpha_1'} \dline \tpl{s'_2,\alpha_2'}
      \end{array}
\end{equation*}

The synchronous product ($\dwavy$) is based on shared actions, defined below.
\begin{equation*}\label{RG PS}\begin{array}{c}
      \begin{array}{l}
        \exists ~e_1\in\alpha_1 \cdot \overline{e_1} = s_1 \xrightarrow{a} s_1' \\
        \exists ~e_2\in\alpha_2 \cdot \overline{e_2} = s_2 \xrightarrow{a} s_2'   
      \end{array}
       \!\land\!
      \begin{array}{l}
      \alpha_1' = \big( \alpha_1 {\cup} \mathsf{on}(e_1,\alpha_1) {\cup} \Gamma^{\oplus}(e_1) \big)
                  \backslash
                  \big( \mathsf{off}(e_1,\alpha_1) {\cup} \Gamma^{\ominus}(e_1)\big)
                  \\
      \alpha_2' = \big( \alpha_2 {\cup} \mathsf{on}(e_2,\alpha_2) {\cup} \Gamma^{\oplus}(e_2) \big)
                  \backslash
                  \big( \mathsf{off}(e_2,\alpha_2) {\cup} \Gamma^{\ominus}(e_2)\big)
      \end{array}
      \\\hline
      \tpl{s_1,\alpha_1} \dwavy \tpl{s_2,\alpha_2} ~ \xrightarrow{a}  ~ 
      \tpl{s'_1,\alpha_1'} \dwavy
              \tpl{s'_2,\alpha_2'}
  \end{array}
  \end{equation*}

This product supports the modelling, e.g., of self-adaptive systems with a layer that manages a given system and the actual system being adapted~\cite{DBLP:conf/ifm/PasslerBDTJ23}.

    \section{The \Marge tool}\label{sec:tool}
 \begin{figure}[t!]
            \centering
            \includegraphics[width = \textwidth]{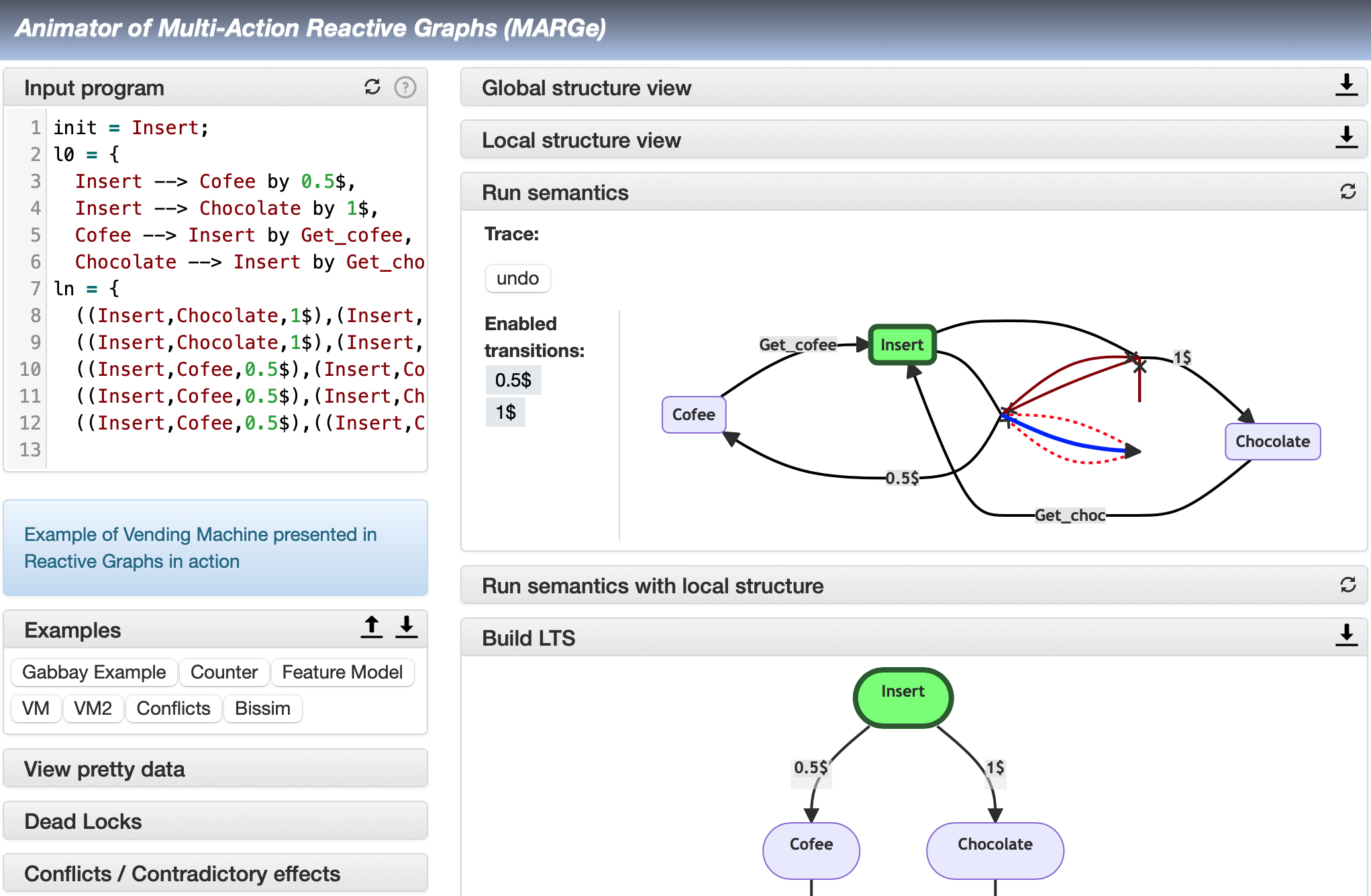}
            \caption{Screenshot of the web interface of the \Marge tool}
            \label{fig:overall-screenshot}
        \end{figure}

This section briefly presents the \Marge tool and its features. A screenshot of its web interface can be found in \cref{fig:overall-screenshot} with the vending machine example from \cref{fig:vd1}.
        \Marge is open-source and developed in \emph{Scala}, using the \emph{Mermaid} library%
        \footnote{Mermaid is  popular markup language for diagrams, cf. \urlpp{https://mermaid.js.org}}
         to produces graphical representations of the reactive graph and its semantics. The tool is compiled to \emph{JavaScript} that is used to build an interactive web page, using the CAOS library~\cite{proenca_caos_2023} which includes support to animate operational semantics and compare semantics.

 \myparagraph{Using the tool.}
        The tool can be used at {\small\url{https://fm-dcc.github.io/MARGe}}, illustrated in \cref{fig:overall-screenshot} with our vending machine example.
        The model is introduced using a textual description in the ``\textsf{Input program}'' widget (top left). It can also be found in the list of examples (middle left). The remaining widgets provide our analysis and visualisations, and can be either collapsed (as the ``\textsf{Global structure view}'') or expanded (as the ``\textsf{Run semanitcs}'').

\myparagraph{Available widgets.} Some of the available widgets are described below.\\[-6mm]
    \begin{itemize}

        \item \textsf{Input program} -- uses a textual notation, mimicking the mathematical structures, not yet optimised to be compact and maintainable.
        
        \item \textsf{Global structure view} -- shows the graphical representation of a reactive graph. A simplified version without hyper edges and deactivated edges is depicted in the widget ``\textsf{local structure view}''.
        
        \item \textsf{Run semantics} -- allows the user to simulate the reactive graph by selecting, at each step, an active transition that should be taken. After selecting this transition the graph is updated, including the active edges and the current state.
        
        \item \textsf{Generated LTS} -- displays the underlying LTS by expanding all possible actions of the the reactive graph (up to a fixed bound).

        \item \textsf{Number of states and edges} -- presents the number of states and edges of both the reactive graph and its encoded LTS. E.g., our vending machine as a similar number of states and edges. But a variation (available online) with a limited stock (instead of limited money) uses 4 states, 5 ground edges, and 3 hyper edges, against 19 states and 20 edges in the encoded LTS.

        \item \textsf{Find strong bisimulation} -- checks if two \mars separated by `$\sim$' in the input program
        are equivalent (i.e., bisimilar), providing either a bisimulation or an an explanation for not finding~one.

        \item \textsf{Conflicts/Contradictory effects} -- finds conflicts when they exist, i.e. traces until a transition that simultaneously tries to activate and deactivate.

        \item \textsf{DeadLocks} -- checks the existence of deadlocks in the behaviour of a given \mar.

        \item \textsf{Products} -- this a set of widgets, presenting the different types of product introduced in this paper.
    \end{itemize}

\section{Conclusions and future work}\label{sec:conclusions}
\Mars can provide a compact and insightful representation of a variety of reconfigurable scenarios, e.g., in the context of communication protocols~\cite{Figueiredo18} and in a biological setting~\cite{SantiagoMF21}.
They are also closely related to van Benthem's game models with adversarial agents~\cite{Benthem05} 
and to Areces et al.'s logic with reconfiguring modal operators~\cite{Areces15}.
Extensions to \mars with the paraconsistency paradigm have also been recently proposed~\cite{CostaFM23}.
Most work on \mars is theoretical, and a small effort has been done to provide tool support and automatization of results.
This paper presents the tool \Marge with basic editor and exploration mechanisms of \mars, including specific analysis such as a search for contradictory effects. 
As future work, we intent to improve the usability of \Marge (e.g., improving the input language), extend it to support fuzzy extensions (to measure, e.g., costs and rewards from applying reconfigurations~\cite{SantiagoMF21,CamposSMF22}), and to integrate a model checker according to a suitable logic. The latter would be similar to how we integrated the mCRL2 toolset~\cite{mcrl2} to analyse connectors~\cite{taming-fsen19} and team automata~\cite{feta}, based on a predecessor of~CAOS~\cite{reolive}.

\subsubsection*{Acknowledgments}
This work is supported by the FCT, the Portuguese funding agency for Science and Technology, with the projects UIDB/04106/2020 (\url{https://doi.org/10.54499/UIDB/04106/2020}), UIDP/04106/2020 (\url{https://doi.org/10.54499/UIDP/04106/2020}), and PTDC/CCI-COM/4280/2021.
It was also supported by the CISTER Research Unit (UIDP/UIDB/\allowbreak04234/2020), financed by National Funds through FCT/MCTES and by project Ibex (ref. PTDC/CCI-COM/4280/2021) financed by national funds through FCT.

  \bibliographystyle{splncs04}
  \bibliography{ref.bib}

\newpage

\appendix

\section{Modelling of a larger example in \Marge}
\label{appendix: FTS}
\begin{figure}
        \centering
        \resizebox{\columnwidth}{!}{%
        \begin{tikzpicture}[node distance={15mm}, thick, main/.style = {draw,rectangle, rounded corners = 2pt}, action/.style = {rectangle, fill = lightgray!50!white, text = black, rounded corners = 2pt}, minimum size = 0.3cm] 
            
            \begin{scope}
                \node[main] (setup) at (0,0){setup};
                \node[main] (ready) at (4,0){ready};
                \node[main] (received) at (8,0) {received};
                \node[main] (routed-safe) at (12,2){routed-safe};
                \node[main] (routed-unsafe) at (12,-2){routed-unsafe};
                \node[main] (sent) at (16,2){sent};
                \node[main] (sent-encrypt) at (16,-2){sent-encrypt};

                \draw[-{Stealth[length=3mm]},out = 70, in = 110, looseness = 15] (setup)  to node[action](safe){safe} (setup);
                \draw[-{Stealth[length=3mm]},out = -70, in = -110, looseness = 15] (setup)  to node[action](unsafe){unsafe} (setup);
                \draw[-{Stealth[length=3mm]},out = 50, in = 130, looseness = 20] (setup)  to node[action](encrypt){encrypt} (setup);
                \draw[-{Stealth[length=3mm]},out = -50, in = -130, looseness = 20] (setup)  to node[action](dencrypt){dencrypt} (setup);
                \draw[-{Stealth[length=3mm]},out = 15, in = 165] (setup)  to node[action]{-} (ready);
                \draw[{Stealth[length=3mm]}-,out = -15, in = -165] (setup)  to node[action]{-} (ready);
                \draw[-{Stealth[length=3mm]}] (ready)  to node[action]{receive} (received);
                \draw[-{Stealth[length=3mm]},out=90,in=180] (received)  to node[action](rsafe){route}(routed-safe);
                \draw[-{Stealth[length=3mm]},out=-90,in=180] (received)  to node[action](runsafe){route}(routed-unsafe);
                \draw[-{Stealth[length=3mm]}] (routed-safe)  to node[action]{send}(sent);
                \draw[-{Stealth[length=3mm]}] (routed-unsafe)  to node[action](sencrypt){send}(sent-encrypt);
                \draw[-{Stealth[length=3mm]},out=90,in=-90] (routed-unsafe)  to node[action](sdencrypt){send} (sent);
                \draw[-{Stealth[length=3mm]},out=90,in=90] (sent)  to node[action]{ready}(ready);
                \draw[-{Stealth[length=3mm]},out=-90,in=-90] (sent-encrypt)  to node[action]{ready} (ready);
            \end{scope}

            \begin{scope}[on background layer]
                \draw[color = red,-{Stealth[length=3mm]},bend left= 1cm] (safe) to (rsafe); 
                \draw[color = red,-{Rays[length=10pt,width=10pt,line width= 1.5pt]},bend left= 1cm] (safe) to (runsafe); 
                \draw[color = red,-{Rays[length=10pt,width=10pt,line width= 1.5pt]},bend left= 1cm,out = -90, in = -90] (unsafe) to (rsafe); 
                \draw[color = red,-{Stealth[length=3mm]},bend left= 1cm,out = -90, in = -90] (unsafe) to (runsafe); 
                \draw[color = red,-{Stealth[length=3mm]},bend left= 1cm] (encrypt) to (sencrypt); 
                \draw[color = red,-{Rays[length=10pt,width=10pt,line width= 1.5pt]},bend left= 1cm] (encrypt) to (sdencrypt); 
                \draw[color = red,-{Rays[length=10pt,width=10pt,line width= 1.5pt]},bend left= 1cm,out = -90, in = -90] (dencrypt) to (sencrypt); 
                \draw[color = red,-{Stealth[length=3mm]},bend left= 1cm,out = -90, in = -90] (dencrypt) to (sdencrypt); 
            \end{scope}
        \end{tikzpicture} 
        }
        \caption{Example adapted from \cite{cordy2013model}}
        \label{fig: FM tikz}
    \end{figure}
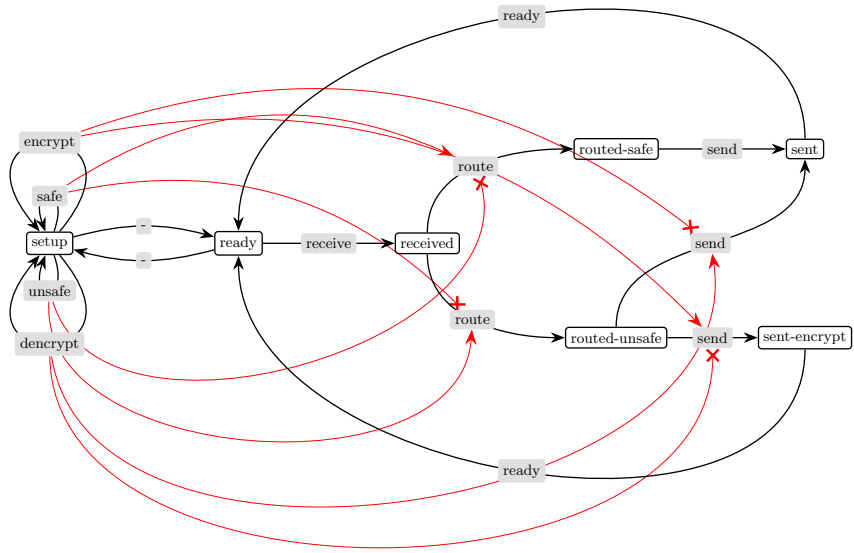

    This sections presents an example describing a feature transitions system (FTS), borrowed from Cordy et al.~\cite[Fig.~1]{cordy2013model}.
   As in the original example, it uses a set of features that can be selected in an initial setup step, triggering the activation and deactivation of edges. Our reactive graph version of this FTS is depicted in \cref{fig: FM tikz}. Variations of this model for \emph{reconfigurable} FTS can also be made, where the feature selection can be modified outside the setup step.

\begin{figure}
            \centering
            \includegraphics[width = \textwidth]{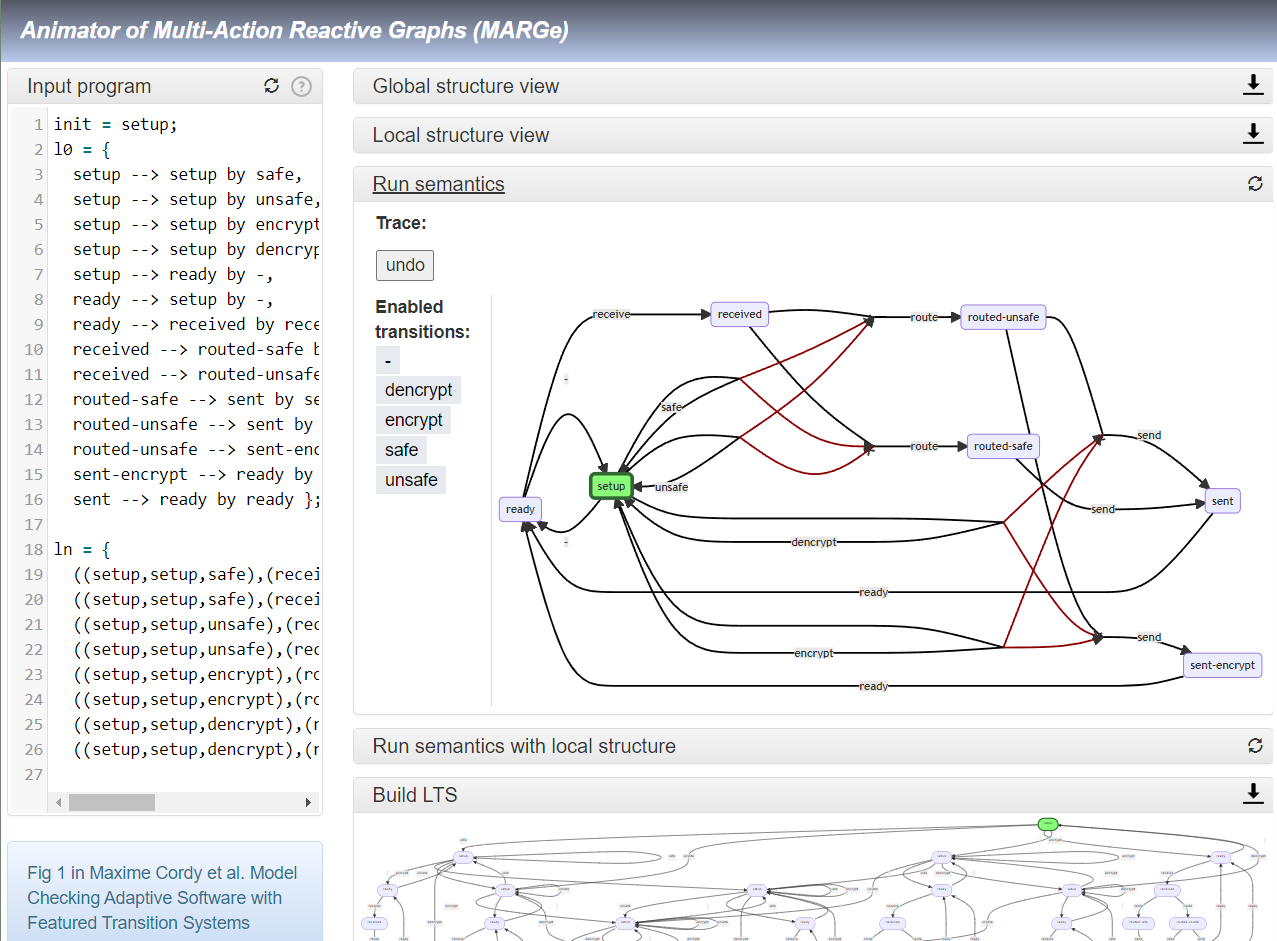}
            \caption{Screenshot of \Marge modelling a FTS}
            \label{fig:FMTool}
        \end{figure}

    A screenshot of \Marge analysing this reactive graph is presented in \cref{fig:FMTool}.

\begin{figure}
            \centering
            \includegraphics[width = .9\textwidth]{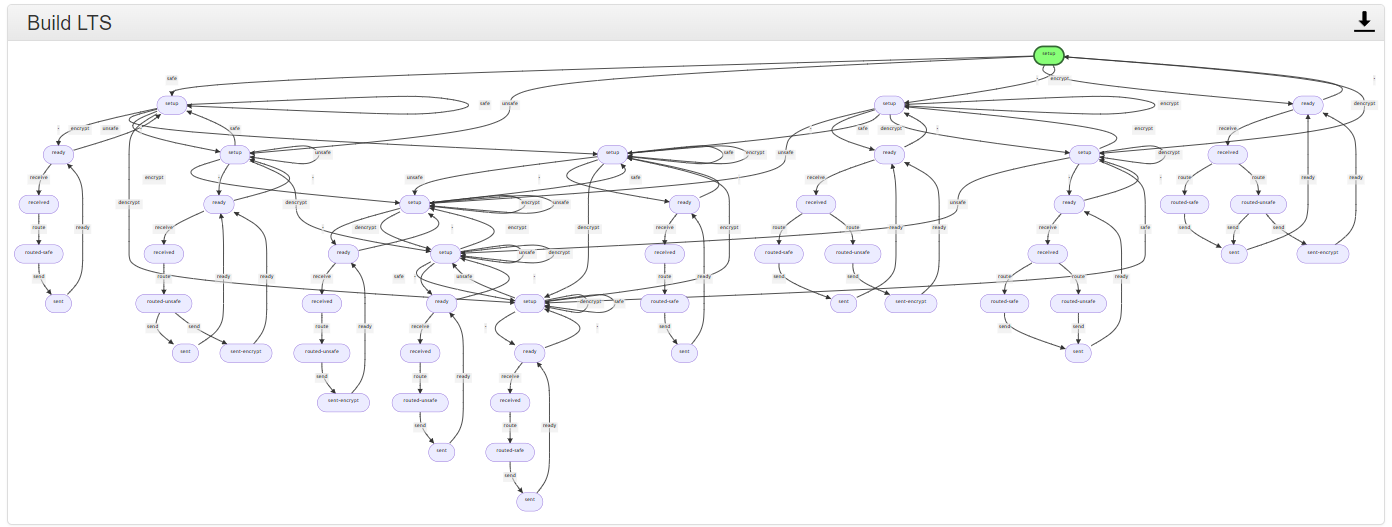}
            \caption{Screenshot of the \textsf{Build LTS} widget with the LTS of the FTS example}
            \label{fig: FMLTS}
        \end{figure}

\begin{figure}
            \centering
            \includegraphics[width = 0.5\textwidth]{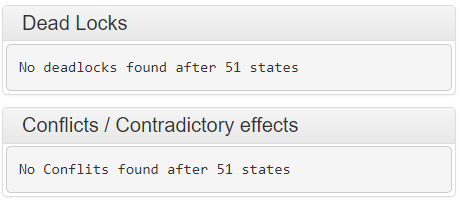}
            \caption{Screenshot of the widgets that search for deadlocks and contradictory effects}
            \label{fig: FMDLC}
        \end{figure}

        In \cref{fig: FMLTS} we can see the explosion of states and the how $\mar$s can compactly represent this systems.
        The reactive graph has 7 states, 14 ground edges, and 8 hyper edges, while the encoded LTS has 51 states and 101 edges.
        One can confirm that this system has no deadlocks and no contradictory effects, using the widgets depicted in \cref{fig: FMDLC}.
        This means that is always possible go to another state, regardless of the current state, and there is no transition that can trigger contradictory effects. 
        
\end{document}